\documentclass[twocolumn,longbibliography,pra,showpacs]{revtex4-1}

\usepackage{amsmath}
\usepackage{amssymb}
\usepackage{xcolor}
\usepackage{graphicx}
\usepackage{bbm}
\usepackage{maybemath}
\usepackage{feynmp-auto}
\usepackage{tikz}
\usetikzlibrary{arrows,decorations.pathmorphing}

\usepackage{natbib}
\usepackage{dcolumn}

\usepackage[%
colorlinks=true,
urlcolor=blue,
linkcolor=blue,
citecolor=blue
]{hyperref}

\def\ii{{\mathrm{i}}}

\makeatletter

\newcommand{\Rmnum}[1]{\expandafter\@slowromancap\romannumeral #1@}

\newcolumntype{.}{D{x}{}{-1}}

\definecolor{duelferred}{rgb}{0.7, 0.2, 0.1}

\definecolor{kellygreen}{rgb}{0.3, 0.73, 0.09}

\definecolor{garrosgreen}{rgb}{0.1, 0.4, 0.1}
\definecolor{dartmouthgreen}{rgb}{0.05, 0.5, 0.06}

\definecolor{duelferred}{rgb}{0.7, 0.2, 0.1}
\definecolor{cambridgeblue}{rgb}{0.1, 0.3, 1.0}
\definecolor{oxfordblue}{rgb}{0.05, 0.2, 0.7}

\definecolor{gold}{rgb}{0.85,.66,0}

\begin{document}

\newcommand{\addrROLLA}{Department of Physics,
Missouri University of Science and Technology,
Rolla, Missouri 65409, USA}

\title{Relativistic and Radiative Corrections to the Dynamic Stark Shift:\\
Gauge Invariance and Transition Currents in the Velocity Gauge}

\author{U. D. Jentschura}
\affiliation{\addrROLLA}
   
\author{C. M. Adhikari}
\affiliation{\addrROLLA}

\begin{abstract}
We investigate the gauge invariance of the 
dynamic (ac) Stark shift under ``hybrid'' gauge 
transformations from the ``length''
($\vec E \cdot \vec r$) to the ``velocity"
($\vec A \cdot \vec p$) gauge.
By a ``hybrid'' gauge transformation,
we understand a transformation in which the 
scalar and vector potentials are modified,
but the wave function remains unaltered.
The gauge invariance of the leading term 
is well known, while 
we here show that gauge invariance 
under perturbations holds only if one takes into 
account an additional correction to the 
transition current, which persists only in the 
velocity gauge.
We find a general expression for this current,
and apply the formalism to radiative and 
relativistic corrections to the dynamic Stark effect,
which is described by the sum of two polarizability 
matrix elements.
\end{abstract}

\pacs{12.20.Ds, 32.80.Rm, 32.70.Cs, 11.15.Bt, 31.15.xp}

\maketitle

% \tableofcontents
%% \newpage

%
% Introduction
%
\section{Introduction}
\label{intro}

One might think that all conceivable questions 
regarding the gauge invariance of physical processes
in quantum electrodynamics (QED) have already been 
addressed in the literature. That is not 
the case. The point is that strictly speaking, a transformation 
from the  ``length'' ($\vec E \cdot \vec r$) to the ``velocity"
($\vec A \cdot \vec p$) gauge requires a gauge
transformation of the wave function, which is, however,
inconvenient to implement in practice, and whose 
necessity is almost always ignored in 
practical calculations~\cite{ScBeBeSc1984}. 
Indeed, a particularly interesting gauge transformation is 
the Power--Zienau transformation, which transforms the QED
Hamiltonian from the $\vec A \cdot \vec p$ (``velocity'') to the 
$\vec E \cdot \vec r$ (``length'') form~\cite{PoZi1959,Pa2004}. 

The question then is which gauge 
should be used in the analysis, e.g., of 
spectroscopic experiments as one models the 
excitation dynamics~\cite{La1952,ScBeBeSc1984,HaEtAl2006}.
In a now famous remark 
on p.~268 of Ref.~\cite{La1952}, Lamb notices that 
the interpretation of the wave function is only 
preserved in the length gauge, and that this gauge 
should be used, therefore, in the description
of his experiments. Specifically, this is because the 
momentum operator retains its physical interpretation
only in the length gauge, without being modified by 
the presence of a nonvanishing vector 
potential, which otherwise makes it necessary to 
distinguish kinetic and canonical momenta~\cite{ScBeBeSc1984}.

Here, we would like to refer to a gauge transformation which 
ignores the phase of the wave function as a ``hybrid'' gauge 
transformation. 
Recently, it has been shown in Ref.~\cite{Je2016} that,
under ``hybrid'' gauge transformations,
two-photon transition matrix elements  are manifestly 
gauge-``dependent'' 
(not gauge invariant) off resonance (i.e., when one
transforms from the length to the velocity gauge
and ignores the gauge transformation of the wave function).
Specifically, in two-photon transitions, the gauge invariance of 
transition matrix elements 
under the hybrid scheme holds only at exact resonance~\cite{Je2016}. 

In contrast, it is well 
known~\cite{Sa1967Adv,Je2004rad,HaEtAl2006,JeLaDKPa2015,JePa2015epjd1}
that a number of other processes which involve
laser-atom interactions, such as the ac Stark shift,
or radiative corrections to the real and imaginary 
part of the polarizability~\cite{JeLaDKPa2015,JePa2015epjd1}, are in fact gauge invariant
under the ``hybrid'' transformations.
The common picture here is that one could, in principle,
formulate these effects in terms of an adiabatic switching of
the interaction Hamiltonian with a factor $\exp(- \varepsilon | t |)$,
where $\varepsilon$ is infinitesimal and $t$ is the time variable,
invoke the Gell--Mann Low theorem [Eq.~(21) of Ref.~\cite{HaJeKe2006}],
and carry out the gauge transformation  of the wave
function at $t = \pm\infty$, where it amounts to the 
identity transformation (because the perturbing fields vanish). 
All processes which allow for such a description, 
have been found to be gauge invariant
under ``hybrid'' 
transformations~\cite{Sa1967Adv,Je2004rad,HaEtAl2006,JeLaDKPa2015,JePa2015epjd1}.

We here investigate questions related to
processes which are gauge invariant 
under ``hybrid'' gauge transformations.
Let us suppose that the (Schr\"{o}dinger) 
Hamiltonian $H$ of the system is being perturbed 
by an additional Hamiltonian $\delta H$.
This perturbation induces a change in 
the energy by $\delta E = \langle \phi | \delta H | \phi \rangle$,
and the wave function perturbation 
is  $| \delta \phi \rangle = 
[1/(E-H)'] \, \delta H | \phi \rangle$,
where $[1/(E-H)']$ is the reduced Green function.
The question we pose is as follows: Which perturbation 
to the interaction Hamiltonian (i.e., to the transition current) 
needs to be added in the 
velocity gauge, for general $\delta H$, in order to ensure gauge invariance 
of energy shifts, when we consider the transformation 
from the velocity to the length gauge?

We shall investigate this question, using the 
ac Stark shift as an example.
Indeed, quite recently, the ac Stark shift has been investigated
in strong laser fields~\cite{JeEvHaKe2003,JeKe2004aop},
with an emphasis on the dressed-state formalism,
and on the nontrivial additional QED
corrections which influence the Mollow spectrum of the 
emitted radiation, beyond the trivial shift of the 
unperturbed atomic levels, due to QED effects.
The relativistic and radiative corrections 
to the incoherent radiation spectrum emitted by the dressed states,
have been analyzed.
By contrast, in a weak laser field,
the atom-laser interaction can be treated perturbatively.
The perturbative effect of a time varying electric field
is commonly referred to as the dynamic or ac 
(``alternating current'') Stark shift~\cite{Ya2003,HaJeKe2006}. 

We organize this paper as follows. 
After recalling fundamental aspects of a gauge transformations
in Sec.~\ref{gauge}, we present in Sec.~\ref{acnr}
a short orientation on the leading-order dynamic (ac) Stark shift. 
In Secs.~\ref{pertlength} and \ref{pertvelo},
we examine the question of how a perturbative potential 
modifies the dynamic polarizability 
and, hence, the ac Stark shift in the 
length and in the velocity gauges, respectively.
A proof of the gauge invariance of the dynamic polarizability 
induced by a perturbative potential is presented in 
Sec.~\ref{proof}.
Two special cases of perturbative potentials are of
phenomenological relevance (see Sec.~\ref{pertrel-and-pertrad}),
namely, \textit{(i)} an effective Lamb-shift potential which describes 
the leading radiative correction to the ac Stark shift,
and  \textit{(ii)} the Hamiltonian describing the 
leading relativistic correction.

%
% Foundations
% 
\section{Foundations}

%
% Leading-order (nonrelativistic) dynamic Stark shift
%
\subsection{Gauge Transformation}
\label{gauge}

We recall that
under an electromagnetic $U(1)$ gauge transformation, 
a wave function $\phi(\vec{r},t)$ transforms as follows,
\begin{subequations}
\begin{align}
\phi(\vec{r},t) \rightarrow \phi'(\vec{r},t)
=\exp\left(\frac{\mathrm{i}e\,\Lambda(\vec{r},t)}{\hbar}\right)\phi(\vec{r},t) \,,
\end{align} 
and the scalar and vector potentials transform as 
\begin{align}
\vec{A}(\vec{r},t)\rightarrow \vec{A\,}'(\vec{r},t)
=\vec{A}(\vec{r},t)+ \vec{\Delta} \Lambda(\vec{r},t) \,,
\\
\Phi(\vec{r},t)\rightarrow \Phi'(\vec{r},t)
=\Phi(\vec{r},t)-\frac{\partial}{\partial t} \Lambda(\vec{r},t)\,,
\end{align}
\end{subequations}
where $\Lambda(\vec{r},t)$ is an arbitrary function of $\vec{r}$ and $t$,
while $\vec{A}(\vec{r},t)$ and  $\Phi(\vec{r},t)$ are, respectively,
the vector and the  scalar potentials.
Under a full gauge transformation of the wave function and the 
potentials, transition matrix elements and energy shifts are 
invariant. However, it is sometimes computationally 
cumbersome to implement a gauge transformation
of both the wave function and potentials,
and one often resorts to a
hybrid gauge-transformation~\cite{BaFoQu1977,Ko1978prl,ScBeBeSc1984,Je2016},
where the wave function is left invariant, 
and only the (vector) potentials are transformed.

%
% Leading-order (nonrelativistic) dynamic Stark shift
%
\subsection{Leading (Nonrelativistic) Dynamic Stark Shift}
\label{acnr}

We assume an atom to be irradiated by a laser with polarization vector
$\hat{\epsilon}_{\rm L}$. To good approximation, 
one may ignore the magnetic 
field which leads to a small perturbation of the interaction.
We implicitly assume that the  atom is in a standing-wave laser field
at a point of maximum electric 
field intensity, where the magnetic field completely vanishes.
This approximation was also made in Ref.~\cite{Ya2003}.
Field-configuration dependent corrections are discussed
in Sec.~IV of Ref.~\cite{HaEtAl2006} and in Sec.~III of Ref.~\cite{AdKaJe2016}.

The dynamic Stark shift $\Delta E_{\rm ac}$ is given by 
\begin{subequations}
\label{lengthP}
\begin{align}
\Delta E_{\rm ac} =& \;
- \frac{e^2 \, I_L \, Q}{2\, c\, \epsilon_0 \, \omega^2} \,,
\\[2ex]
Q =& \; \omega^2 \,
\left( \left< \phi \left| (\vec{\epsilon}_{\rm L} \cdot \vec{x}) \, 
\frac{1}{H - E + \omega} \, (\vec{\epsilon}_{\rm L} \cdot \vec{x}) 
\right| \phi \right> \right.\nonumber\\[2ex]
& \; +\left. \left< \phi \left| (\vec{\epsilon}_{\rm L} \cdot \vec{x}) \,
\frac{1}{H - E - \omega} \, (\vec{\epsilon}_{\rm L} \cdot \vec{x})
\right| \phi \right> \right)\,.
\end{align}
\end{subequations}
Here $\omega$ is the angular laser frequency,
and $I_L$ is the laser intensity.
Here and in the following,
we will assume, without loss of generality, that the 
laser field is oriented along the $z$-axis,
i.e.,~$\vec{\epsilon}_{\rm L} = \hat{z}$.
The corresponding canonically conjugate momentum 
will be denoted by $p^z = - {\rm i}\, \partial/(\partial z)$.
We can restrict the 
discussion to a $z$-polarized laser field
with frequency $\omega$
because the only atomic states under investigation 
here are $S$ states which are isotropic.
In contrast, the dynamic Stark shift 
would depend on the magnetic quantum 
number of $P$ states and states with higher orbital 
angular momenta. 

%
% Perturbations
%
\section{Perturbations}

%
% Length gauge 
%
\subsection{Length--Gauge Perturbation}
\label{pertlength}

In the following, we use natural units with 
$\epsilon_0 = \hbar = c = 1$, as is customary in the treatment
of relativistic corrections in atomic physics.
Thus, for example, in our unit system, the Rydberg constant $R_\infty$
is equal to $\alpha^2 m/2$. Our unit of length is
the reduced electron Compton wavelength.
We consider a perturbation to the
dynamic Stark shift~(\ref{lengthP}) due to some perturbation
$\delta H$ which is added to the 
Schr\"{o}dinger Hamiltonian. Because
both relativistic as well as the leading logarithmic radiative corrections
can be expressed in terms of perturbative potentials,
the formalism developed here allows for a 
unified treatment of the relativistic and 
radiative corrections to the dynamic polarizability,
as discussed below in Sec.~\ref{pertrel-and-pertrad}. 

In the length gauge, 
the dynamic Stark shift is proportional to the quantity $Q$
[see Eq.~(\ref{lengthP})] which may be expressed as
\begin{equation}
\label{acstarklength}
Q = \omega^2 \, \rho\,,
\qquad 
\rho = \rho_1 + \rho_2\,,
\end{equation}
where in turn (the reference state is $| \phi \rangle$),
\begin{subequations}
\begin{align}
\label{defrho}
\rho_1 = \left< \phi \left| z \, 
\frac{m}{H - E + \omega} \,
z \right| \phi \right> \,,\\
\rho_2 = \left< \phi \left| z \, 
\frac{m}{H - E - \omega} \,
z \right| \phi \right> \,.
\end{align}
\end{subequations}
We now consider the first-order perturbation
received by the quantity $\rho$ via the action of 
a perturbative Hamiltonian $\delta H$ which modifies the 
Schr\"odinger Hamiltonian $H$ according to 
$H \to H + \delta H$. The perturbation $\delta H$ leads to a 
perturbation of the energy of the bound state,
of the wave function and, of course, $\delta H$ 
also constitutes a correction to the Hamiltonian $H$ in the propagator 
denominator. In general, we have the following first-order perturbations:
\begin{subequations}
\begin{align}
\label{perth}
H & \to H + \delta H\,,\\[2ex]
\label{perten}
E & \to E + \delta E \,, \qquad
\delta E = \left< \phi | \delta H | \phi \right>\,, \\[2ex]
\label{pertwave}
| \phi \rangle & \to | \phi \rangle +
\left. \left. \left( \frac{1}{E - H} \right)' \, \delta H 
\right| \phi \right>\,.
\end{align}
\end{subequations}
Here, the prime in the operator $1/\left(E - H \right)'$ 
indicates that the reference state is excluded from
the spectral decomposition of the operator
(``reduced Green function'').
The correction received by $Q$ via the action of $\delta H$ 
is then
\begin{equation}
\label{corrlength}
\delta Q = \omega^2 \, \delta \rho \,,
\end{equation}
where $\delta \rho$ is the sum of six terms,
\begin{equation}
\delta \rho = \sum_{j=1}^6 \delta \rho_j\,.
\end{equation}
Here, $\delta \rho_1$ and $\delta \rho_2$ are perturbations
to the Hamiltonian,
\begin{subequations}
\label{defrho:all6}
\begin{align}
\label{defDeltarho1}
\delta \rho_1 &= -\left< \phi \left| z 
\frac{m}{H - E + \omega}  \delta H 
\frac{1}{H - E + \omega} 
z \right| \phi \right>\,,
\\[2ex]
\label{defDeltarho2}
\delta \rho_2 &= -\left< \phi \left| z 
\frac{m}{H - E - \omega}  \delta H 
\frac{1}{H - E - \omega}
z \right| \phi \right>\,.
\end{align}
The quantities $\delta \rho_3$ and $\delta \rho_4$ are energy perturbations,
\begin{align}
\label{defDeltarho3}
\delta \rho_3 &= \left< \phi \left| z \, 
\left( \frac{m}{H - E + \omega} \right)^2 \,
z \right| \phi \right>\, 
\frac{\left< \phi \left| \delta H \right| \phi \right>}{m}\, ,
\\[2ex]
\label{defDeltarho4}
\delta \rho_4 &= \left< \phi \left| z \, 
\left( \frac{m}{H - E - \omega} \right)^2 \,
z \right| \phi \right>\, 
\frac{\left< \phi \left| \delta H \right| \phi \right>}{m}\, .
\end{align}
Finally, the terms $\delta \rho_{5,6}$ are perturbations to the wave function, 
\begin{align}
\label{defDeltarho5}
\delta \rho_5 &= 2\,\left< \phi \left| z 
\frac{m}{H - E + \omega} 
z  \left( \frac{1}{E - H} \right)' \delta H
\right| \phi \right>\,,
\\[2ex]
\label{defDeltarho6}
\delta \rho_6 &= 2\,\left< \phi \left| z 
\frac{m}{H - E - \omega} 
z \left( \frac{1}{E - H} \right)' \delta H
\right| \phi \right>\,.
\end{align}
\end{subequations}

%
% FORMULATION OF THE PROBLEM 
%
\subsection{Velocity--Gauge Perturbation}
\label{pertvelo}

The dynamic Stark shift,
in the velocity gauge,
is proportional to the quantity $Q'$
which may be expressed as
\begin{equation}
\label{acstarkvel}
Q' = \chi \,, \qquad 
\chi = \chi_1 + \chi_2 + \chi_3\,,
\end{equation}
where
\begin{subequations}
\label{defchi}
\begin{eqnarray}
\label{defchi1}
\chi_1 &=& \left< \phi \left| \frac{p^z}{m} \, 
\frac{m}{H - E + \omega} \,
\frac{p^z}{m} \right| \phi \right> \,,
\\[2ex]
\label{defchi2}
\chi_2 &=& \left< \phi \left| \frac{p^z}{m} \, 
\frac{m}{H - E - \omega} \,
\frac{p^z}{m} \right| \phi \right> \,,
\\[2ex]
\label{defchi3}
\chi_3 &=& - \left< \phi | \phi \right> = -1 \,.
\end{eqnarray}
\end{subequations}
The seagull term is responsible for $\chi_3$.
The prime in $Q'$ denotes the velocity-gauge form of the
correction. It is instructive to observe that 
the large-$\omega$ 
asymptotic of $Q'$ read as follows,
\begin{align}
\label{largeomega}
Q' & = -1 -
2\,\frac{m}{\omega^2} \, \left< \phi \left| \frac{p^z}{m} \, 
(H - E) \,
\frac{p^z}{m} \right| \phi \right> \nonumber\\[2ex]
&= -1 -
\frac{m}{\omega^2} \, 
\left< \phi \left| \frac13\,
\vec{\nabla}^2(V) \right| \phi \right> 
\nonumber\\[2ex]
& = -1 -
\frac{m}{\omega^2} \, 
\left< \phi \left| \frac43\, 
\frac{\pi (Z\alpha)}{m^2} \, 
\delta^{(3)}(\vec{r})\,
\right| \phi \right> \nonumber\\[2ex]
&= - 1 - \frac43\, \frac{(Z\alpha)^4}{n^3} \,
\frac{m^2}{\omega^2} \, \delta_{\ell 0}\,,
\end{align}
where we assume a hydrogenic state 
with principal quantum number $n$ that is nonvanishing at
the origin only for $S$ symmetry,
\begin{equation}
\left< \phi \left| \delta^{(3)}(\vec{r}) \right| \phi \right> =
\frac{(Z \alpha m)^3}{\pi n^3} \, \delta_{\ell 0} \,.
\end{equation}
The first-order correction to the dynamic polarizability,
in the velocity gauge, is
\begin{equation}
\label{dyncorrvel}
\delta Q' = \delta \chi\,,
\end{equation}
where again the prime denotes the velocity-gauge form of the 
correction. 
Eventually, we desire to show that $\delta Q = \delta Q'$.
Just like its length-gauge counterpart $\delta \rho$,
the velocity-gauge correction $\delta \chi$ is the sum of 
various terms, 
\begin{equation}
\label{deltachi}
\delta \chi = \sum_{j=1}^8 \delta \chi_j\,.
\end{equation}
Here, $\delta \chi_1$ and $\delta \chi_2$ are perturbations
of the Hamiltonian,
\begin{subequations}
\label{defDeltachi:all8}
\begin{align}
\label{defDeltachi1}
\delta \chi_1 &= - \left< \phi \left| \frac{p^z}{m} \, 
\frac{m}{H - E + \omega} \,
\delta H \,
\frac{1}{H - E + \omega} \,
\frac{p^z}{m} \right| \phi \right>\,,
\\[2ex]
\label{defDeltachi2}
\delta \chi_2 &= - \left< \phi \left| \frac{p^z}{m} \, 
\frac{m}{H - E - \omega} \,
\delta H \,
\frac{1}{H - E - \omega} \,
\frac{p^z}{m} \right| \phi \right>\,.
\end{align}
The quantities $\delta \chi_3$ and $\delta \chi_4$ are energy perturbations,
\begin{align}
\label{defDeltachi3}
\delta \chi_3 &= \left< \phi \left| \frac{p^z}{m} \, 
\left( \frac{m}{H - E + \omega} \right)^2 \,
\frac{p^z}{m} \right| \phi \right>\, 
\frac{\left< \phi \left| \delta H \right| \phi \right>}{m}\,,
\\[2ex]
\label{defDeltachi4}
\delta \chi_4 &= \left< \phi \left| \frac{p^z}{m} \, 
\left( \frac{m}{H - E - \omega} \right)^2 \,
\frac{p^z}{m} \right| \phi \right>\, 
\frac{\left< \phi \left| \delta H \right| \phi \right>}{m}\, .
\end{align}
The terms $\delta \chi_{5,6}$ are perturbations to the 
wave function, 
\begin{align}
\label{defDeltachi5}
\delta \chi_5 &= 2\,\left< \phi \left| \frac{p^z}{m} \, 
\frac{m}{H - E + \omega} \,
\frac{p^z}{m} \, \left( \frac{1}{E - H} \right)' \delta H
\right| \phi \right>\,,
\\[2ex]
\label{defDeltachi6}
\delta \chi_6 &= 2\,\left< \phi \left| \frac{p^z}{m} \, 
\frac{m}{H - E - \omega} \,
\frac{p^z}{m} \, \left( \frac{1}{E - H} \right)' \delta H
\right| \phi \right>\,.
\end{align}
Quite surprisingly, in the velocity gauge,
there are two more terms, 
\begin{align}
\label{defDeltachi7}
\delta \chi_7 =\; & 
2\,{\rm i}\,\left< \phi \left| \frac{p^z}{m} \,
\frac{m}{H - E + \omega} \,
[\delta H, z] \right| \phi \right>\,, 
\\[2ex]
\label{defDeltachi8}
\delta \chi_8 =\; &
2\,{\rm i}\,\left< \phi \left| \frac{p^z}{m} \,
\frac{m}{H - E - \omega} \,
[\delta H, z] \right| \phi \right>\,. 
\end{align}
\end{subequations}
These corrections are due to a modification 
of the transition current in the velocity gauge,
\begin{equation}
\label{delta_j}
\frac{p^i}{m} \to \frac{p^i}{m} + \delta j^i \,,
\qquad
\delta j^i = \ii \, [ \delta H, x^i ] \,,
\end{equation}
with the correction $\delta j^i$ perturbing both transition currents
in the polarizability matrix element.

For clarification, we should add that
the correction to the wavefunction 
(\ref{pertwave}) is orthogonal to the 
first-order wave function (conservation of the norm),
and hence, the seagull-term contribution $\chi_3$ 
receives no correction
due to the perturbative potential
[see Eq.~(\ref{defchi3})].

%
% Gauge Invariance
%
\section{Gauge Invariance}

%
% Proof of gauge invariance 
%
\subsection{Proof of Gauge Invariance}
\label{proof}

First, let us point out that 
the gauge invariance of the {\em leading-order} dynamic polarizability 
[Eq.~(\ref{acstarklength}) vs.~(\ref{acstarkvel})]
requires the relation
\begin{equation}
\label{aceasy}
Q' = \chi = \omega^2 \, \rho = Q \,.
\end{equation}
We will skip the details of the 
derivation of this identity which may be found in~\cite{Sa1967Adv}
and on pp.~357 -- 359 of Ref.~\cite{CTDRGr1989}. Indeed,
the verification 
of the identity $Q' = Q$ is a rather easy, albeit somewhat 
tedious exercise involving the repeated application of 
the commutator relation
\begin{equation}
\frac{p^z}{m} = {\rm i}\, [H, z] =
{\rm i}\, [H - E \pm \omega, z] \,.
\end{equation}
The gauge invariance $Q = Q'$ 
of the leading-order dynamic Stark shift~(\ref{aceasy})
raises pertinent questions concerning a 
potentially similar 
relation $\delta Q = \delta Q'$ for the 
first-order correction to this quantity.
In detail, for the nonrelativistic case, the gauge-invariance relation is
\begin{equation}
\label{dyninvariance}
Q = Q' \; \Leftrightarrow \;
\omega^2 \, \left( \sum_{i=1}^2 \rho_i \right) =
\sum_{i=1}^3 \chi_i\,,
\end{equation}
with two terms in the length gauge, but three terms in 
the velocity gauge.  For the correction,
the appropriate form is  
\begin{align}
\label{master}
\delta Q = \delta Q' \; \Leftrightarrow \;
\omega^2 \, \left( \sum_{i=1}^6 \delta \rho_i \right) = \; &
\sum_{i=1}^8 \delta \chi_i\,.
\end{align}

We now present the derivation of the 
formula (\ref{master}) (gauge invariance of the 
correction to the dynamic polarizability 
mediated by a perturbative potential $\delta H$),
by first investigating the
velocity-gauge form of the correction, and then
transforming into the length gauge.
For $\delta\chi_1$ as defined in (\ref{defDeltachi1}), we have
\begin{subequations}
\begin{align}
\label{Deltachi1tolength}
\delta \chi_1& =-\left< \phi \left| \frac{p^z}{m} \, 
\frac{m}{H - E + \omega} \,
\delta H\,
\frac{1}{H - E + \omega} \,
\frac{p^z}{m} \right| \phi \right> 
\nonumber\\[2ex]
&= -\omega^2  \,
\left< \phi \left| z \, 
\frac{m}{H - E + \omega} \,
\delta H\,
\frac{1}{H - E + \omega} \,
z \right| \phi \right> \nonumber\\[2ex]
&\; + 2 \, \omega  \left< \phi \left| z \, \frac{m}{H - E + \omega}
\delta H\,
z \right| \phi \right> + 
\left< \phi \left| z  m\,\delta H\, z \right| \phi \right>
\nonumber\\[2ex]
&= \omega^2 \, \delta \rho_1 + 2 \, \omega \, 
\left< \phi \left| z \, \frac{m}{H - E + \omega} \, \delta H\,
z \right| \phi \right>\nonumber\\[2ex]
&\; + \left< \phi \left| z \, m\,\delta H\, z \right| \phi \right>\,.
\end{align}
An analogous relation, valid for $\delta\chi_2$,
can be obtained by the replacement
$\omega \to - \omega$ in Eq.~(\ref{Deltachi1tolength}).
We transform $\delta\chi_3$ as defined in (\ref{defDeltachi3})
according to 
\begin{align}
\label{Deltachi3tolength}
\delta \chi_3 &= \left< \phi \left| \frac{p^z}{m} \, 
\left( \frac{m}{H - E + \omega} \right)^2 \,
\frac{p^z}{m} \right| \phi \right> \, 
\frac{\left< \phi | \delta H | \phi \right>}{m} 
\nonumber\\[2ex]
=\; & \omega^2 \, \delta \rho_3 - 2 \, \omega \, 
\left< \phi \left| z \, \frac{m}{H - E + \omega} \,
z \right| \phi \right> \, \left< \phi | \delta H | \phi \right> \nonumber\\[2ex]
&+ \left< \phi \left| z^2 \right| \phi \right> \, 
\left< \phi | m\,\delta H | \phi \right> \,.
\end{align}
Again, an analogous relation, valid for $\delta\chi_4$,
can be obtained by the replacement 
$\omega \to - \omega$ in Eq.~(\ref{Deltachi3tolength}).
For $\delta \chi_5$, the following relation is useful,
\begin{align}
\label{Deltachi5tolength}
\delta \chi_5 = \;&
2\,\left< \phi \left| \frac{p^z}{m} \, 
\frac{m}{H - E + \omega} \,
\frac{p^z}{m} \, \left( \frac{1}{E - H} \right)' \delta H
\right| \phi \right> 
\nonumber\\[2ex]
=\;& \omega^2\,\delta \rho_5
- 2 \omega \, \left< \phi \left| z \, \frac{m}{H - E + \omega} \,
z \, \delta H \right| \phi \right> \nonumber\\[2ex]
&+ 2 \omega \, \left< \phi \left| z \, \frac{m}{H - E + \omega} \,
z \right| \phi \right> \, \left< \phi \left| \delta H \right| \phi \right>
\nonumber\\[2ex]
& + \left< \phi \left| z^2 m\, \delta H \right| \phi \right> -
\left< \phi \left| z^2 \right| \phi \right>\,
\left< \phi | m\,\delta H | \phi \right>\nonumber\\[2ex]
&+ 2 \omega \, \left< \phi \left| z^2 
\left( \frac{1}{E - H} \right)' m\,\delta H \right| \phi \right>\,.
\end{align}
\end{subequations}
Replacement of $\omega$ by $-\omega$ in 
Eq.~(\ref{Deltachi5tolength})  yields $\delta \chi_6$.
%
%
% In the last transformation (\ref{Deltachi5tolength}),
% we have used
% %
% \begin{align}
% \label{whow}
% & \left< \phi \left| z \, (H - E) \, z \, 
% \left( \frac{1}{E - H} \right)'\, \delta H \right| \phi \right> =
% \nonumber\\
% & = \frac12 \, 
% \left< \phi \left| [ z , H - E ] \, z \, 
% \left( \frac{1}{E - H} \right)'\, \delta H \right| \phi \right> +
% \frac12 \, 
% \left< \phi \left| z \, [ H - E , z] \, 
% \left( \frac{1}{E - H} \right)'\, \delta H \right| \phi \right> 
% \nonumber\\
% & \qquad +
% \frac12 \, 
% \left< \phi \left| z^2  \, (H - E) \, 
% \left( \frac{1}{E - H} \right)'\, \delta H \right| \phi \right> 
% \nonumber\\
% & =
% \frac{{\rm i}\,m}{2} \, 
%  \left< \phi \left| (p^z \, z - z \, p^z) \,  
% \left( \frac{1}{E - H} \right)'\, \delta H \right| \phi \right> 
% - \frac12 \, 
% \left< \phi \left| z^2 \, (1 - | \phi \rangle\,\langle \phi|)\,
% \delta H \right| \phi \right> 
% \nonumber\\
% & = \frac12 \, \left< \phi \left| z^2 \right| \phi \right> \,
% \left< \phi \left| \delta H \right| \phi \right> -
% \frac12 \, \left< \phi \left| z^2 \, \delta H \right| \phi \right> \,.
% \nonumber\\
% \end{align}
%
Using Eqs.~(\ref{Deltachi1tolength})---(\ref{Deltachi5tolength}),
we finally obtain the simple and compact relation
\begin{align}
\label{dyninvariance2}
\omega^2  \sum_{i=1}^6 \delta \rho_i  =  &
\sum_{i=1}^6 \delta \chi_i 
- 2 \, \omega \, 
\left< \phi \left| z \, \frac{m}{H - E + \omega} 
\left[ \delta H, z \right] \right| \phi \right> 
\nonumber\\[2ex]
& + 2 \, \omega \, 
\left< \phi \left| z \, \frac{m}{H - E - \omega} \, 
\left[ \delta H, z \right] \right| \phi \right> \nonumber\\[2ex]
&- 2\,\left< \phi \left| z \, m\,\left[ \delta H, z \right] \right| \phi \right>\,.
\end{align}
We recall that the expression $\sum_{i=1}^6 \delta \chi_i$
represents the sum of the wave-function correction, the 
correction to the Hamiltonian, and the correction due to the 
energy perturbation mediated by a perturbative potential $\delta H$.

What remains to be shown is that the sum of the additional
terms $\delta \chi_7$ and $\delta \chi_8$,
as defined in Eqs.~(\ref{defDeltachi7}) and~(\ref{defDeltachi8}),
reproduces the remaining terms on the right-hand side of
Eq.~(\ref{dyninvariance2}). This can be accomplished as follows,
\begin{align}
\sum^8_{i=7} \delta \chi_i 
=\; &  2\,{\rm i}\,\Big\{ \Big< \phi \Big| \frac{p^z}{m} \, 
\frac{m}{H - E + \omega} \,
[\delta H, z] \Big| \phi \Big>\nonumber\\[2ex]
& +\Big< \phi \Big| \frac{p^z}{m} \, 
\frac{m}{H - E - \omega} \,
[\delta H, z] \Big| \phi \Big> \Big\}
\nonumber\\[2ex]
=&  -2\Big\{ \Big< \phi \Big| [H - E + \omega, z] 
\frac{m}{H - E + \omega} 
[\delta H, z] \Big| \phi \Big> \nonumber\\[2ex]
&+ (\omega \to -\omega)  \Big\}
\nonumber\\[2ex]
=\;& - 2 \, \omega \, 
\left< \phi \left| z \, \frac{m}{H - E + \omega} \, 
\left[ \delta H, z \right] \right| \phi \right> 
\nonumber\\[2ex]
& + 2 \, \omega \, 
\left< \phi \left| z \, \frac{m}{H - E - \omega} \, 
\left[ \delta H, z \right] \right| \phi \right> \nonumber\\[2ex]
&- 2 \,
\left< \phi \left| z \, m\,\left[ \delta H, z \right] \right| \phi \right>\,.
\end{align}
We recognize, in the last line, the terms on the right-hand side of
Eq.~(\ref{dyninvariance2}). This concludes the proof of 
Eq.~(\ref{master}). 

%
% Figure 1
%
\begin{figure}[t!]%
\begin{minipage}{0.99\linewidth}
\begin{center}
\begin{center}\includegraphics[width=0.99\linewidth]{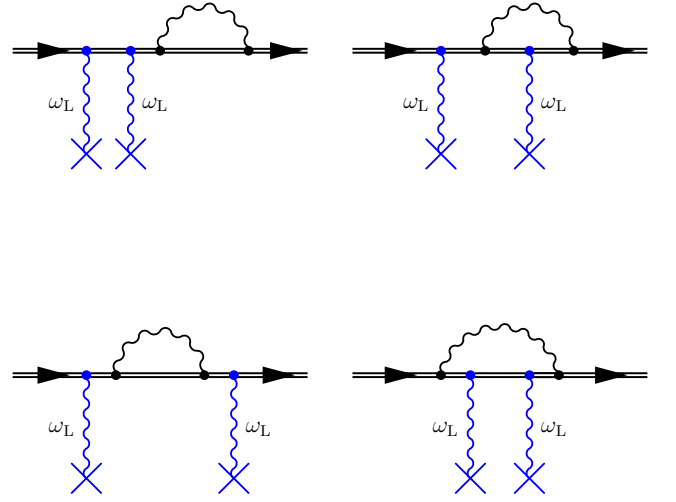}\end{center}%
\caption{\label{fig1}%
(color online).
Feynman diagrams for the self-energy radiative correction
to the dynamic Stark shift. Interactions with the
external laser field are labeled with $\omega_{\rm L}$.}%
\end{center}
\end{minipage}
\end{figure}

%
% Two important special cases
%
\subsection{Relativistic and Radiative Effects}
\label{pertrel-and-pertrad}

%
% Leading relativistic correction
%
\subsubsection{Leading relativistic correction}

A Foldy--Wouthuysen transformation of the Dirac--Coulomb 
Hamiltonian~\cite{BjDr1964}
gives us the following relativistic correction (see e.g.~\cite[p.~19]{Je2003dipl}) 
\begin{equation}
\label{deltaH}
\delta H = 
-\frac{\vec{p\;}^4}{8m^3} + 
\frac{\pi (Z\alpha)}{2 m^2} \, \delta^{(3)}(\vec{r}) +
\frac{Z\alpha}{4 m^2 r^3}\,
\vec{L} \cdot \vec{S} \,.
\end{equation}
For the relativistic correction to the current,
we need the commutator
\begin{equation}
[ \delta H, z] = -\frac{1}{8m^3} \, [\vec{p\;}^4, z] =
\frac{\rm i}{2 m^3} \, p^z \, \vec{p\;}^2 \,.
\end{equation} 
The two additional terms, in this case
[see Eqs.~(\ref{defDeltachi7}) and~(\ref{defDeltachi8})], are
\begin{align}
\label{defa1}
\delta \chi_{\delta H,7} 
=& 2\,\left< \phi \left| \frac{p^z}{m} \, 
\frac{m}{H - E + \omega} \,
\left( - \frac{1}{2 m^3} \,p^z\,\vec{p\;}^2 \right) 
\right| \phi \right>\,,
\\[2ex]
\label{defa2}
\delta \chi_{\delta H,8} 
=& 2\,\left< \phi \left| \frac{p^z}{m} \, 
\frac{m}{H - E - \omega} \,
\left( - \frac{1}{2 m^3} \,p^z\,\vec{p\;}^2 \right) 
\right| \phi \right>\,.
\end{align}
We observe that these terms are exactly equal to the  terms
on the right-hand side of Eq.~(\ref{dyninvariance2}),
which leads us immediately to 
the gauge-invariance relation
\begin{equation}
\omega^2 \, \sum_{i = 1}^6 \delta \rho_i(\delta H)
= \sum_{j = 1}^8 \delta \chi_j(\delta H) \,.
\end{equation}
The two additional terms $\delta \chi_7$ and 
$\delta \chi_8$ in the velocity gauge are definitely necessary
in order to ensure gauge invariance;
they are due to correction 
to the current which prevails only in the velocity, but not in the 
length gauge (see p.~21 of Ref.~\cite{Je2003dipl}).

%
% Leading radiative correction
%
\subsubsection{Leading radiative correction}

%Normalized local potential,

Inspired by effective field-theory, or nonrelativistic 
quantum electrodynamics~\cite{CaLe1986}, 
we here pursue an effective treatment in which the 
leading logarithmic QED correction due to radiative photons is described by
an effective Lamb-shift potential (see also Fig.~\ref{fig1})
\begin{equation}
\delta H =
\delta V_{\rm Lamb} = \frac43\,\alpha\,(Z\alpha)\, \ln[(Z\alpha)^{-2}]\,
\frac{\delta^{(3)}(\vec{r})}{m^2}\,.
\end{equation}
It is sometimes useful to consider a ``standard'' 
perturbative potential~\cite{Je2003jpa}
\begin{equation}
\label{standard}
\delta V = \frac{\pi (Z\alpha)}{m^2} \, \delta^{(3)}(\vec{r})\,,
\end{equation}
which is related to $\delta V_{\rm Lamb}$ by a simple prefactor,
\begin{align}
\label{VLamb}
\delta V_{\rm Lamb}& = \frac{4 \alpha}{3 \pi} \,
[\pi (Z\alpha)] \,\ln[(Z\alpha)^{-2}] \, \frac{\delta^{(3)}(\vec{r})}{m^2}\nonumber\\[2ex]
&= \frac{4 \alpha}{3 \pi} \, \ln[(Z\alpha)^{-2}] \, \delta V\,.
\end{align}

The standard potential (\ref{standard})
leads to a ``normalized'' energy shift with unit prefactors,
\begin{equation}
\delta E (\phi \ell_j) = \frac{(Z\alpha)^4\,m}{n^3}\, \delta_{\ell 0}\,,
\end{equation}
for hydrogenic states with the principal quantum number $n$,
orbital quantum number $\ell$, and total angular momentum
quantum number $j$.
If a numerical evaluation is desired, then the 
radiative corrections $\delta Q$ can be read off
from the sum of the various terms listed in Eq.~(\ref{defrho:all6}).
A generalization to the leading effect of vacuum polarization,
replacing $\delta V$ by the Uehling potential~\cite{Ue1935}
is immediate.

In general, for a perturbative potential $\delta V$
that fulfills $[\delta V, z] = 0$, the additional
terms $\delta \chi_7$ and $\delta \chi_8$ are not necessary.
In this case,
the gauge-invariance statement can be summarized as follows,
\begin{equation}
[\delta V, z] = 0 \;\;\Rightarrow \;\; 
\omega^2 \sum_{i = 1}^6 \delta \rho_i(\delta V) =
\sum_{j=1}^6 \delta \chi_j(\delta V)\,,
\end{equation}
leaving out $\delta \chi_7 = \delta \chi_8 = 0$.

%
% Figure 2
%
\begin{figure}[t!]%
\begin{minipage}{0.99\linewidth}
\begin{center}
\begin{center}\includegraphics[width=0.99\linewidth]{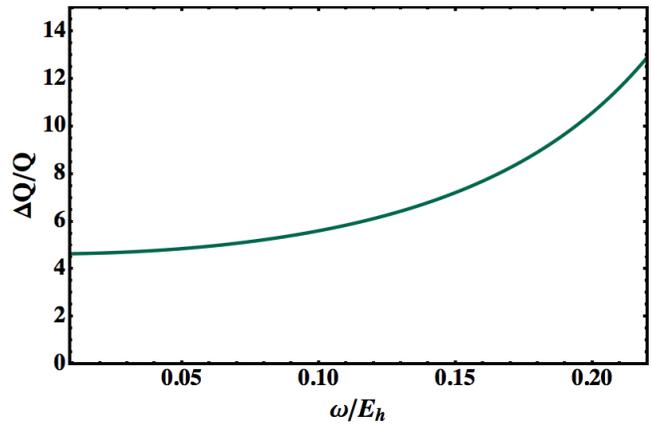}\end{center}%
\caption{\label{fig2}%
(color online). Ratio of the first-order radiative correction of the dynamic polarizability to 
the unperturbed dynamic polarizability as a function of the laser photon energy $\omega$
(we set $\hbar = 1$), divided by the Hartree energy $E_h$. 
The data are obtained for the ground state of hydrogen. 
The quantity $\Delta Q$ is defined in Eq.~\eqref{DeltaQ}.}%
\end{center}
\end{minipage}
\end{figure}

%
% Figure 3
%
\begin{figure}[t!]%
\begin{minipage}{0.99\linewidth}
\begin{center}
\begin{center}\includegraphics[width=0.99\linewidth]{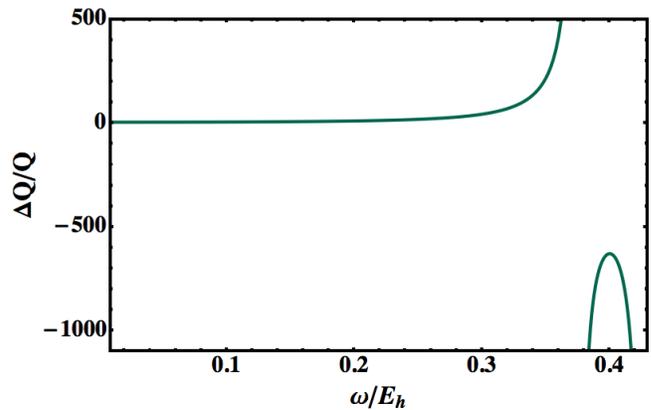}\end{center}%
\caption{\label{fig3}%
(color online). Same as Fig.~\ref{fig2}, but in a frequency range 
which covers the intermediate $2P$ state, where the laser frequency 
can excite the $1S$---$2P$ transition resonantly.
The plot is included for reference. Of course, 
near resonance, the second-order perturbation treatment
of the atom-laser interaction, which is the basis for 
Eq.~\eqref{lengthP}, breaks down, and the dressed-state formalism has to 
be used (see Ref.~\cite{JeKe2004aop}).
The $2P$ resonance is responsible for the first peak in the radiative correction
at $\omega = \tfrac38 \, E_h$, and the second pole is due to 
the zero of the unperturbed matrix element $Q$ at $\omega = 0.429538 \, E_h$.
All calculations are performed in the non-recoil approximation.
The figure illustrates the dramatic increase of the radiative correction 
as the resonance is approached.}
\end{center}
\end{minipage}
\end{figure}

In Figs.~\ref{fig2} and~\ref{fig3}, we present numerical data for 
the frequency-dependent radiative correction
(the ``logarithmic coefficient'')
\begin{equation}
\label{DeltaQ}
\frac{\Delta Q}{Q} =
\left( \frac{4 \alpha^3}{3 \pi} \, \ln( \alpha^{-2} ) \right)^{-1} \, 
\frac{\delta Q}{Q} \,,
\end{equation}
where $\delta Q$ is the leading logarithmic radiative correction 
due to the effective potential~\eqref{VLamb},
evaluated for the ground-state of hydrogen. 
The numerical calculations use techniques originally developed
in self-energy calculations~\cite{JePa1996}.
Large coefficients are obtained for the leading logarithmic correction.

%
% Conclusions 
%
\section{Conclusions}

We have investigated the gauge invariance of the 
dynamic (ac) Stark shift under the ``hybrid'' gauge 
transformation from the length to the 
velocity gauge. The length-gauge perturbations 
due to a perturbative Hamiltonian $\delta H$ have been 
discussed in Sec.~\ref{pertlength}, 
while the velocity-gauge formulation is 
given in Sec.~\ref{pertvelo}.

In the velocity gauge, six perturbations,
two each to the Hamiltonian, to the energy and to the wave function,
have been given in Eq.~\eqref{defrho:all6},
while the eight terms in the velocity gauge 
can be found in Eq.~\eqref{defDeltachi:all8}.
Gauge invariance amounts to showing the 
identity~\eqref{master}.
This is accomplished in Sec.~\ref{proof},
where we also give a general form of the 
additional correction to the current,
which is necessary to include in the 
velocity gauge [see Eq.~\eqref{delta_j}].
Indeed, the general form of the correction to the current,
induced by the perturbative Hamiltonian $\delta H$,
\begin{equation}
\delta j^i = \ii \, [ \delta H, x^i ] \,,
\end{equation}
has not been recorded in the literature, 
to the best of our knowledge, and constitutes
a main result of our investigations.

While all derivations discussed in the current paper have been given 
for one-electron atoms, the generalization
to many-electron systems is straightforward:
One simply sums over the electron coordinates.
One should add that the derivation here is related to the one
recently presented in Appendix A of~\cite{Je2004rad}
in the context of the gauge invariance of radiative
corrections to the two-photon decay width,
and to Ref.~\cite{JeLaDKPa2015,JePa2015epjd1} for 
the gauge invariance of 
the imaginary part of the atomic polarizability.

%%% The attentive reader will have noticed that 
%%% we have neglected the magnetic interaction
%%% This is valid provided we assume, with Ref.~\cite{Ya2003},
%%% a standing dipole laser field,
%%% with the atom being positioned at the 
%%% maximum of the electric field. 
%%% In this case, there will be no magnetic field present 
%%% and no retardation
%%% effects given by the changing phase of the 
%%% laser field as a function of the atomic position.

In general, the length gauge is favorable
for the formulation of relativistic corrections
because the number of terms is smaller in this 
gauge, and the interactions are formulated in terms
of gauge-invariant field strengths ($\vec E$ and $\vec B$)
instead of gauge-dependent scalar and vector 
potentials ($\Phi$ and $\vec A$);
see Refs.~\cite{ScBeBeSc1984,Je2016} for further discussions 
on this point. 

%% This is one of the reasons why recently, 
%% an effective field-theory known as long-wavelength 
%% quantum electrodynamics has been formulated in 
%% the length gauge~\cite{Pa2004}, rather than 
%% the velocity gauge.

A specific picture is emerging from the recent 
investigations on gauge invariance:
For resonant processes which involve eigenstates
of the same energy, of the combined atom$+$radiation-field system,
the ``hybrid'' gauge invariance holds.
This is, e.g., the case for the two-photon decay width~\cite{Je2004rad}, 
where the initial $2S$ state has the same energy 
as the final state (atom is in the $1S$ state,
and two photons are in the radiation field).
This is also the case for two-photon transition 
matrix elements, provided the final state has the same
energy as the initial state, plus the energy 
of the two absorbed photons (i.e., at resonance, see 
Refs.~\cite{Je2004rad,Je2016}).
For the dynamic polarizability studied in the current article, 
the resonance condition is always 
met because relevant matrix elements describe the absorption
of a laser photon and the concomitant emission of 
that same photon.
So, the initial state considered in our investigations
here has the same 
energy as the final atomic state, which is in fact identical 
to the initial state (it has the same number of laser photons,
and the same atomic state).

The deeper reason for the ``hybrid''
gauge invariance of resonant processes
lies in the possibility of formulating 
such energy perturbations in terms of adiabatically 
switched fields and potentials; the 
gauge transformation of the initial and final states
of the wave function at $t \to \pm \infty$ amounts
to the identity transformation because
the adiabatically switched fields and potentials 
vanish in that same limit.
Hence, the gauge transformation of the wave function
can be omitted.
This general picture is confirmed by the investigations presented here,
and augmented by the general form of the transition current 
which has to be added in the velocity gauge.

%
% Acknowledgments
%
\acknowledgments{This research has been supported by the
National Science Foundation (Grant PHY-1710856)
and by the Missouri Research Board. }

\end{document}